\documentclass[10pt,journal]{IEEEtran}

\usepackage{amsmath}
\usepackage{amssymb}
\usepackage[dvips]{graphicx}
\usepackage{setspace}
\usepackage{epsfig}

\newcommand{\K}{{\sf{K}}}

\newcommand{\n}{\mathbf{n}}

\newcommand{\I}{\mathbf{I}}
\newcommand{\s}{\mathbf{s}}
\newcommand{\E}{\mathcal{E}}

\newcommand{\C}{{\sf{C}}}

\newcommand{\e}{\epsilon}

\newcommand{\rr}{{\mathbf{r}}}

\newcommand{\figsize}{0.48}

\newcommand{\tsnr}{{\text{\footnotesize{SNR}}}}

\newcommand{\ssnr}{\text{\scriptsize{SNR}}}

\newtheorem{theo:awgnderivatives}{Theorem}
\newtheorem{theo:cfderivatives}[theo:awgnderivatives]{Theorem}
\newtheorem{theo:ncfderivatives}[theo:awgnderivatives]{Theorem}
\newtheorem{theo:awgnderivfsk}[theo:awgnderivatives]{Theorem}
\newtheorem{theo:fskbitenergyasymp}[theo:awgnderivatives]{Theorem}
\newtheorem{theo:ncfskbitenergyasymp}[theo:awgnderivatives]{Theorem}
\newtheorem{theo:oofskasympt}[theo:awgnderivatives]{Theorem}
\newtheorem{cor:finiteParSlope}[theo:awgnderivatives]{Theorem}

\newtheorem{corr:awgnasympt}{Corollary}
\newtheorem{corr:awgnbitenergy}[corr:awgnasympt]{Corollary}
\newtheorem{corr:awgnM3bitenergy}[corr:awgnasympt]{Corollary}
\newtheorem{corr:ncfadingasympt}[corr:awgnasympt]{Corollary}
\newtheorem{corr:ncfadingbitenergy}[corr:awgnasympt]{Corollary}
\newtheorem{corr:cfderivatives}[corr:awgnasympt]{Corollary}
\newtheorem{corr:fadingfsk}[corr:awgnasympt]{Corollary}
\newtheorem{corr:oofskhard}[corr:awgnasympt]{Corollary}

\begin{document}

\title{The Impact of Hard-Decision Detection on the Energy Efficiency of Phase and Frequency
Modulation}



%
\author{\authorblockN{Mustafa Cenk Gursoy}
\thanks{Mustafa Cenk Gursoy is with the Department of Electrical
Engineering, University of Nebraska-Lincoln, Lincoln, NE 68588
(e-mail : gursoy@engr.unl.edu). }
\thanks{This work was supported in part by the NSF
CAREER Grant CCF-0546384. The material in this paper was presented
in part at the IEEE International Symposium on Information Theory
(ISIT), Nice, France, in June 2007, and at the IEEE International
Symposium on Information Theory (ISIT), Toronto, Canada, in July
2008.}
}


\maketitle

\begin{abstract}
The central design challenge in next generation wireless systems is to have these systems operate at high bandwidths and provide high data rates while being cognizant of the energy consumption levels especially in mobile applications. Since communicating at very high data rates prohibits obtaining high bit resolutions from the analog-to-digital (A/D) converters, analysis of the energy efficiency under the assumption of hard-decision detection is called for to accurately predict the performance levels. In this paper, transmission over the additive white
Gaussian noise (AWGN) channel, and coherent and noncoherent fading
channels is considered, and the impact of hard-decision detection on
the energy efficiency of phase and frequency modulations is
investigated. Energy efficiency is analyzed by studying the capacity of these modulation schemes and the energy required to send one bit of information reliably in the low signal-to-noise ratio ($\tsnr$) regime. The capacity of hard-decision-detected phase and frequency modulations is characterized at low $\tsnr$ levels through closed-form expressions for the first and second derivatives of the capacity at zero $\tsnr$. Subsequently, bit energy requirements in the low-$\tsnr$ regime are identified.
%
%
The
increases in the bit energy incurred by hard-decision detection and channel
fading are quantified. Moreover, practical design guidelines for the
selection of the constellation size are drawn from the analysis of
the spectral efficiency--bit energy tradeoff. 
\\


\end{abstract}

\section{Introduction} \label{sec:intro}

Energy efficiency is of paramount importance in many communication
systems and particularly in mobile wireless systems due to the
scarcity of energy resources. Energy efficiency can be measured by
the energy required to send one information bit reliably. It is
well-known that for Gaussian channels subject to average power
constraints, the minimum received bit energy normalized by the noise
spectral level is $\frac{E_b}{N_0}_{{\min}} = -1.59$ dB regardless
of the availability of channel side information (CSI) at the
receiver (see e.g., \cite{Golay} -- \cite{Lapidoth Shamai}, and \cite{Verdu}). Golay
\cite{Golay} showed that this minimum bit energy can be achieved in
the additive white Gaussian noise (AWGN) channel by pulse position
modulation (PPM) with vanishing duty cycle when the receiver employs
threshold detection. Indeed, Turin \cite{Turin} proved that any
orthogonal $M$-ary modulation scheme with envelope detection at the
receiver achieves the normalized bit energy of $-1.59$ dB in the
AWGN channel  as $M \to \infty$. It is further shown in
\cite{Jacobs} and \cite{Pierce} that $M$-ary orthogonal
frequency-shift keying (FSK) achieves this minimum bit energy
asymptotically as $M \to \infty$ also in noncoherent fading channels
where neither the receiver nor the transmitter knows the fading
coefficients. These studies demonstrate the asymptotical high energy
efficiency of orthogonal signaling even when the receiver performs
hard-decision detection. As also well-known by now in the digital
communications literature \cite{Proakis}, these results are shown by
proving that the error probabilities of orthogonal signaling can be
made arbitrarily small as $M \to \infty$ as long as the normalized
bit energy (or equivalently $\tsnr$ per bit) is greater than $-1.59$
dB. As indicated by the unbounded growth of $M$, the minimum bit
energy is in general achieved at infinite bandwidth or equivalently
as the spectral efficiency (rate in bits per second divided by
bandwidth in Hertz) goes to zero.

Indeed for average power limited channels, the bit energy required
for reliable communication decreases monotonically with increasing
bandwidth \cite{Verdu_cost}, \cite{Verdu}. This is the fundamental
bandwidth-power tradeoff. Recently, Verd\'u \cite{Verdu} has offered
a more subtle analysis of the tradeoff of bit energy versus spectral
efficiency. In this work, the wideband slope, which is the slope of
the spectral efficiency curve at zero spectral efficiency, has
emerged as a new analysis tool to measure energy and bandwidth
efficiency in the low-power regime. It is shown that if the receiver
has perfect knowledge of the fading coefficients, quaternary
phase-shift keying (QPSK) is an optimally efficient modulation
scheme achieving both the minimum bit energy of $-1.59$ dB and the
optimal wideband slope in the low $\tsnr$ regime. This indicates
that besides orthogonal signaling, phase modulation is also
well-suited for energy efficient operation. However, it should be
noted that asymptotic efficiency of QPSK holds under the assumption
that the receiver performs soft detection. Verd\'u \cite{Verdu} has
also provided expressions for the minimum bit energy and wideband
slope of the quantized QPSK. We note that phase modulation is a widely used technique for information
transmission, and the performance of coded phase modulation has been
of interest in the research community since the 1960s. One of the
early works was conducted in \cite{Wyner} where the capacity and
error exponents of a continuous-phase modulated system, in which the
transmitted phase can assume any value in $[-\pi,\pi)$, is studied.
More recent studies include \cite{Pierce}, \cite{Kramer},
and \cite{Kaplan}--\cite{Zhang}.

As discussed above, high energy efficiency requires operation in the
wideband regime in which the spectral efficiencies are low. This is
is achieved by either decreasing the data rates or increasing the
bandwidth. If the system has large bandwidth, then the data rates
are high. For instance, if the total signal power is $P = 1$ mW and
the bandwidth is $B = 1$ GHz, then the capacity of the AWGN channel
is $C = B \log_2 \left(1 + \frac{P}{N_0 B}\right) \approx 27.9$
Gbits/s\footnote{We have assumed that $N_0 = 4\times 10^{-21}$ W/Hz
\cite{ProakisSalehi}.}. If the bandwidth is increased to $B = 10$
GHz, the capacity becomes 245.7 Gbits/s. Similarly, high rates are
also achieved in fading channels when the available bandwidth is
large. For instance, in current practical applications, wideband
CDMA and ultrawideband systems offer high data rate services by
using large bandwidths \cite{Garg}. Additionally, operating at high
bandwidths and providing high data rates while conserving the energy
in mobile applications are the key features of next generation
wireless systems which have the goal of offering mobile multimedia
access. For instance, one of the features of fourth generation (4G)
systems will be the ability to support multimedia services at low
transmission cost \cite[Chap. 23, available online]{Garg}. On the
other hand, at these very high transmission rates, obtaining high
bit resolutions from A/D converters may either be not possible or
prohibitively expensive. Therefore, in such cases, the performance
of soft detection will be a loose upper bound on the actual system
performance, and analysis under the assumption of hard-decision
detection will provide more faithful results. Moreover, even if the
data rates are not high, hard-decision detection of the received
signals
might be preferred when reduction in the computational burden is
required \cite{Proakis}. Such a requirement, for instance, may be
enforced in sensor networks that consist of low-cost, low-power,
small-sized sensor nodes \cite{Luo}. Therefore, it is timely and
practically relevant to study the energy efficiency of phase and
frequency modulations in the wideband
regime when the receiver performs hard-decision detection. 

The contributions of this paper are the following:
\begin{enumerate}

\item We obtain closed-form expressions for the first and second derivatives at zero $\tsnr$ of the hard-decision-detected PSK
capacity for arbitrary modulation size $M$.

\item We find
the bit energy required at zero spectral efficiency and wideband
slope when PSK is employed at the transmitter. The analysis is initially performed for
noncoherent fading channels, and subsequently specialized to the AWGN and coherent fading channels. We quantify the increase in the bit energy requirements due to hard-decision detection and channel fading.
\item We study the energy efficiency of
hard-decision-detected on-off frequency-shift keying (OOFSK) modulation
which is a general orthogonal signaling scheme that combines orthogonal
FSK and on-off keying (OOK) and introduces peakedness
in both time and frequency. We show that the bit energy
requirements grow without bound with decreasing $\tsnr$ if the
peakedness in both time and frequency is limited.
We identify the impact upon the energy efficiency of
the number of orthogonal frequencies, $M$, and the duty cycle of OOK.
We prove a sufficient condition on how fast the duty cycle has to
vanish with decreasing $\tsnr$ in order to
approach the fundamental bit energy limit of $-1.59$dB.
\end{enumerate}

The organization of the rest of the paper is as follows. In Section
\ref{sec:channelmodel}, we describe the channel model. The energy
efficiency of phase modulation is investigated in Section
\ref{sec:psk}. $M$-ary OOFSK modulation  and its special case
$M$-ary FSK modulation are considered in Section \ref{sec:fsk}.
Section \ref{sec:conclusion} includes our conclusions.

\section{Channel Model} \label{sec:channelmodel}

We consider the following channel model 
\begin{gather}
\rr_k = h_k \s_{x_k} + \n_k \quad k = 1,2,3 \ldots
\end{gather}
where $x_k$ is the discrete input, $\s_{x_k}$ is the transmitted
signal when the input is $x_k$, and $\rr_k$ is the received signal
during the the $k^{\text{th}}$ symbol duration. $h_k$ is the channel
gain. $h_k$ is a fixed constant in unfaded AWGN channels, while in
flat fading channels, $h_k$ denotes the fading coefficient.
$\{\n_k\}$ is a sequence of independent and identically distributed
(i.i.d.) zero-mean circularly symmetric Gaussian random vectors with
covariance matrix $E\{\n \n^\dagger\} = N_0 \I$ where $\I$ denotes
the identity matrix.
We assume that the system has an average energy constraint of
$E\{\|\s_{x_k}\|^2\} \ \le \E \quad \forall k$.

At the transmitter, if $M$-ary PSK modulation is employed for
transmission, the discrete input, $x_k$, takes values from $\{1,
\ldots, M\}$. If $x_k = m$, then the transmitted signal in the
$k^{\text{th}}$ symbol duration is
\begin{gather}
s_{x_k} = s_m = \sqrt{\E}e^{j \theta_{m}}
\end{gather}
where $\theta_{m} = \frac{2\pi (m-1)}{M}$ for $ m =1, \ldots, M, $
is one of the $M$ phases available in the constellation. In the case
of PSK modulation, since $s_{x_k}$ is a one-dimensional complex point, we opted to not use the boldface representation. Accordingly,
the output $r_k$ and the noise $n_k$ are one complex-dimensional
points. The receiver is assumed to perform hard-decision detection.
Therefore, each received signal $r_k$ is mapped to one of the points
in the constellation set $\{\sqrt{\E}e^{j2\pi (m-1)/M}, m = 1,
\ldots, M\}$ before going through the decoder. We assume that
maximum likelihood decision rule is used at the detector.

In \cite{gursoyoofsk}, we have introduced the on-off frequency-shift
keying (OOFSK) modulation by overlaying frequency-shift keying (FSK)
on on-off keying (OOK). In $M$-ary OOFSK modulation, the transmitter
either sends no signal with probability $1-\nu$ or sends one of $M$
orthogonal FSK signals each with probability $\nu/M$. Hence,  $\nu
\in (0,1]$ can be seen as the duty cycle of the transmission. In
this case, the discrete input takes values from $x_k \in
\{0,1,2,\ldots, M\}$. If $x_k = 0$, then there is no transmission
and the geometric representation of the transmitted signal is the
$M$-complex dimensional vector $\s_0 = (0,0,\ldots,0)$. On the other
hand, if $x_k = m \neq 0$, an FSK signal is sent and the geometric
representation is given by
\begin{gather}
\s_{x_k} = \s_m = (s_{m,1}, s_{m,2}, \ldots, s_{m,M}) \quad m =
1,2,\ldots, M,
\end{gather}
where $s_{m,m} = \sqrt{\E/\nu}\,e^{j\theta_m}$ and $s_{m,i} = 0$ for
$i \neq m$. The phases $\theta_m$ can be arbitrary. Note that in
$M$-ary OOFSK modulation, we have $M+1$ possible input signals including the no signal transmission. Therefore, no signal transmission being a part of the modulation also conveys a message to the receiver. While FSK signals have energy
$\E/\nu$, the average energy of OOFSK modulation is $\E$. Hence, the
peak-to-average power ratio of signaling is $1/\nu$. In the OOFSK
transmission and reception model, the received signal $\rr_k$ and
noise $\n_k$ are also $M$-dimensional. It is assumed that the
receiver performs energy detection on the received vector $\rr_k$.
Finally, note that OOFSK is a general orthogonal signaling format
and specializes to regular orthogonal FSK if $\nu = 1$, and to OOK
if $M = 1$ and $\nu \neq 1$.


We remark that in both PSK and OOFSK cases, the channel, after
hard-decision detection, can be regarded as a discrete channel with
finitely many inputs and outputs. Henceforth, capacity and achievable rate expressions throughout the paper will be obtained considering these discrete channels.

\section{Energy Efficiency of Phase Modulation} \label{sec:psk}


\subsection{Noncoherent Rician Fading Channels}

In this section, we study the performance of phase modulated signals
when they are hard-decision detected. We initially consider  transmission of PSK signals over  noncoherent Rician channels in which neither receiver nor transmitter knows the fading coefficients. Results for this channel are subsequently
specialized to obtain the performance results of PSK in unfaded AWGN
channels and coherent fading channels. Hence, we first assume that
the fading coefficients $\{h_k\}$, whose realizations are unknown at
the transmitter and receiver due to the noncoherence assumption, are
i.i.d. proper complex Gaussian random variables with mean $E\{h_k\}
= d \neq 0$ and variance $E\{|h_k - d|^2\} = \gamma^2$. We further
assume that the channel statistics, and hence $d$ and $\gamma^2$,
are known both at the transmitter and receiver. Note that $d \neq 0$
is required because phase cannot be used to transmit information in
a noncoherent Rayleigh fading channel where $d = 0$.

In the noncoherent Rician channel model, the conditional probability
density function (pdf) of the channel output given the input
is a conditionally complex Gaussian pdf and is given by\footnote{Since the channel is memoryless, we henceforth, without
loss of generality, drop the time index $k$ in the equations for the
sake of simplification.}
\begin{align}
f_{r|s_m}(r | s_m) =\frac{1}{\pi (\gamma^2|s_m|^2 + N_0)}
\,e^{-\frac{|r - ds_m|^2}{\gamma^2|s_m|^2 + N_0}}. 
\label{eq:nfcondprobr}
\end{align}
Recall that $\{s_m = \sqrt{\E} e^{j\theta_m}\}$ are the PSK signals
and hence $|s_m| = \sqrt{\E}$ for all $m = 1, \ldots, M$. Due to
this constant magnitude property, it can be easily shown that the
maximum likelihood detector selects $s_k$ as the transmitted signal
if\footnote{The decision rule is obtained when we assume, without
loss of generality, that $d = |d|$.}
\begin{gather}\label{eq:ncdecisionrule}
\text{Re}(r s_k^*) > \text{Re}(r s_i^*) \quad \forall i \neq k
\end{gather}
where $s_k^*$ is the complex conjugate of $s_k$, and $\text{Re}()$
denotes the operation that selects the real part. We denote the
signal at the output of the detector by $y$ and assume that $ y \in
\{1, \ldots, M\}. $ Note that $y = l$ for $l = 1, \ldots, M$ means
that the detected signal is $\sqrt{\E} e^{j2\pi (l-1) /M}$. Under
the decision rule (\ref{eq:ncdecisionrule}), the decision region for
$y = l$ is the two-dimensional region
\begin{gather} \label{eq:ncdecisionregions}
D_l = \left\{r = |r| e^{j \theta}:  \frac{(2l-3)\pi}{M} \le \theta <
\frac{(2l-1)\pi}{M} \right\}
\end{gather}
for $l = 1,2, \ldots, M$.
With hard-decision detection at the receiver, the resulting channel
is a symmetric, discrete, memoryless channel with input $x \in \{1,
\ldots, M\}$ and output $y \in \{1, \ldots, M\}$. The transition
probabilities are given by
\begin{align}
P_{l,m}  &= P(y = l | x = m)
\\
&=P\left(\frac{(2l-3)\pi}{M} \le \theta
< \frac{(2l-1)\pi}{M} | x = m\right)
\\
&=
\int_{\frac{(2l-3)\pi}{M}}^\frac{(2l-1)\pi}{M} f_{\theta |
s_m}(\theta | s_m) \,d \theta
\end{align}
where $f_{\theta | s_m}(\theta | s_m)$ is the conditional
probability density function of the phase of the received signal
given that the input is $x = m$, and hence the transmitted signal is
$s_m$. It is well-known that the capacity of this symmetric channel
is achieved by equiprobable inputs and the resulting capacity
expression \cite{Cover} is\footnote{Throughout the paper, $\log$ is
used to denote the logarithm to the base $e$, i.e., the natural
logarithm. Additionally, the subscript ''nc" in $C_{M,nc}$ signifies
the noncoherent channel.}
\begin{align}
C_{M,nc}(\tsnr) &= \log M - H(y|x = 1)
\\
&= \log M + \sum_{l = 1}^{M}
P_{l,1} \log P_{l,1} \label{eq:mcap}
\end{align}
where $\tsnr = \frac{\E}{N_0}$, $H(\cdot)$ is the entropy function,
and $P_{l,1} = P(y = l | x = 1)$. In order to evaluate the capacity
of general $M$-ary PSK transmission with a hard-decision detector,
the transition probabilities
\begin{gather}
P_{l,1} = P(y = l| x = 1) =
\int_{\frac{(2l-3)\pi}{M}}^\frac{(2l-1)\pi}{M} f_{\theta |
s_1}(\theta | s_1) \,d \theta
\end{gather}
should be computed. Starting from (\ref{eq:nfcondprobr}) and noting that the conditional joint magnitude and phase distribution is given by
\begin{gather}
f_{|r|,\theta|s_1}(|r|,\theta | s_1) =\frac{|r|}{\pi (\gamma^2|s_1|^2 + N_0)}
\,e^{-\frac{|r|^2 + |d|^2|s_1|^2-2|r||d||s_1|\cos\theta}{\gamma^2|s_1|^2 + N_0}}
\end{gather}
where, without loss of generality, we have assumed that $d = |d|$, we can
easily find that for $\theta \in [0,2\pi)$, $f_{\theta|s_1}(\theta|s_1)$ is given by \eqref{eq:nfcondprobtheta} on the next page
\begin{figure*}
\begin{align}
f_{\theta|s_1}(\theta|s_1) = &\frac{1}{2 \pi} \,
e^{-\frac{|d|^2\tsnr}{\gamma^2 \tsnr + 1}} +
\sqrt{\frac{|d|^2\tsnr}{\pi(\gamma^2 \tsnr + 1)}} \cos \theta
\,e^{-\frac{|d|^2\tsnr}{\gamma^2 \tsnr + 1} \sin^2 \theta} \left( 1
- Q\left(\sqrt{2\frac{|d|^2\tsnr}{\gamma^2 \tsnr + 1} \cos^2
\theta}\right)\right) \label{eq:nfcondprobtheta}
\end{align}
\hrule
\end{figure*}
where $Q(x) = \int_{x}^\infty \frac{1}{\sqrt{2\pi}} \,e^{-t^2/2} \,
dt$ is the Gaussian $Q$-function\footnote{See also \cite{Kramer} and references therein for a similar formula of the phase probability density function derived for the AWGN channel.}. Since $f_{\theta|s_1}$ is rather
complicated, closed-form capacity expressions in terms of
$Q$-functions are available only for the special cases of $M = 2$
and 4:
\begin{gather}
C_{2, nc}(\tsnr) = \log 2 - h\left(Q\left(\sqrt{\frac{2|d|^2
\tsnr}{\gamma^2 \tsnr + 1}}\right)\right), \text{ and}
\\
C_{4,nc}(\tsnr) = 2C_{2,nc}\left(\frac{\tsnr}{2}\right)
\end{gather}
where $ h(x) = - x \log x - (1-x)\log(1-x)$ is the binary entropy
function. For the other cases, the channel capacity can only be
found through numerical integration and computation.


On the other hand, the behavior of the capacity in the low-$\tsnr$
regime can be accurately predicted through the second-order Taylor
series expansion of the capacity\footnote{It can be easily seen from the smoothness and boundedness of $f_{\theta|s_1}$ in \eqref{eq:nfcondprobtheta} that $C_{M,nc}(\tsnr)$ is continuous and differentiable in $\tsnr$.}: $$C_{M,nc}(\tsnr) =
\dot{C}_{M,nc}(0) \tsnr + \ddot{C}_{M,nc}(0) \frac{\tsnr^2}{2} +
o(\tsnr^2)$$ where $\dot{C}_{M,nc}(0)$ and $\ddot{C}_{M,nc}(0)$
denote the first and second derivatives, respectively, of the
channel capacity (in nats/symbol) with respect to $\tsnr$ at $\tsnr
= 0$. In the following result, we provide closed-form expressions
for these derivatives. Note that the wideband regime in which
$\tsnr$ per unit bandwidth is small can equivalently be regarded as
the low-$\tsnr$ regime.

\begin{theo:ncfderivatives} \label{theo:ncfderivatives}
The first and second derivatives of $C_{M,nc}(\tsnr)$ in nats per
symbol at $\tsnr = 0$ are given by
\begin{align}\label{eq:ncfadingfirstderivative}
\hspace{-.33cm}
\begin{split}
\dot{C}_{M,nc}(0) &= \left\{
\begin{array}{ll}
\frac{2|d|^2}{\pi} & M = 2
\\
\frac{M^2|d|^2}{4\pi} \sin^2 \frac{\pi}{M} & M \ge 3
\end{array} \right.,
\quad \text{and}
\\
\ddot{C}_{M,nc}(0) &= \left\{
\begin{array}{ll}
\frac{8}{3\pi}\left( \frac{1}{\pi} -1\right)|d|^4 - \frac{4 |d|^2
\gamma^2}{\pi} & M = 2
\\
\infty & M = 3
\\
\frac{4}{3\pi}\left( \frac{1}{\pi} -1\right)|d|^4 - \frac{4 |d|^2
\gamma^2}{\pi} & M = 4
\\
\psi(M)|d|^4-\frac{|d|^2 \gamma^2}{2\pi}M^2 \sin^2\frac{\pi}{M} & M
\ge 5
\end{array} \right.
\end{split}
\end{align}
respectively, where
\begin{align}
\psi(M) = \frac{M^2}{16\pi^2} \bigg( &(2-\pi)\sin^2 \frac{2\pi}{M} +
(M^2 - 4\pi)\sin^4 \frac{\pi}{M} \nonumber
\\
&- 2M \sin^2
\frac{\pi}{M}\sin\frac{2\pi}{M}\bigg). \label{eq:psi}
\end{align}
\end{theo:ncfderivatives}

\emph{Proof}: See Appendix \ref{app:proofthm1}.

The derivative expressions in (\ref{eq:ncfadingfirstderivative}) are
in closed-form and can be computed easily. Therefore, the
low-$\tsnr$ approximation of the capacity of $M$-ary PSK can be
readily obtained from $$C_{M,nc}(\tsnr) \approx \dot{C}_{M,nc}(0)
\tsnr + \ddot{C}_{M,nc}(0) \frac{\tsnr^2}{2}.$$ The following
corollary provides the asymptotic behavior as $M \to \infty$. In
this asymptotic regime, the transmitted signal is the continuous
phase which can take any value in $[-\pi, \pi)$.

\begin{corr:ncfadingasympt}
In the limit as $M \to \infty$, the first and second derivatives of
the capacity at zero $\tsnr$ converge to
\begin{gather}
\lim_{M\to \infty} \dot{C}_{M,nc}(0) = \frac{\pi|d|^2}{4} \quad \text{and} \\
 \lim_{M \to \infty} \ddot{C}_{M,nc}(0) =
\frac{(\pi^2-8\pi+8)|d|^4}{16}-\frac{|d|^2 \gamma^2 \pi}{2}.
\end{gather}
\end{corr:ncfadingasympt}

In the low-power regime, the tradeoff between bit energy and
spectral efficiency is a key measure of performance \cite{Verdu}.
The normalized energy per bit can be obtained from $ \frac{E_b}{N_0}
= \frac{\tsnr \log 2}{C(\tsnr)} $ where $C(\tsnr)$ is the channel
capacity in nats/symbol. The maximum achievable spectral efficiency
in bits/s/Hz is given by $ \C\left(\frac{E_b}{N_0} \right) =
C(\tsnr)\log_2e \text{ bits/s/Hz} $ if we, without loss of
generality, assume that one symbol occupies a 1s $\times$ 1Hz
time-frequency slot. Two important notions regarding the
spectral-efficiency/bit-energy tradeoff in the low power regime are
the bit-energy required at zero
spectral efficiency and wideband slope, given by 
\begin{gather}
\left.\frac{E_b}{N_0}\right|_{\C = 0} = \frac{\log2}{\dot{C}(0)},
\text{ and } S_0 = \frac{2 (\dot{C}(0))^2}{-\ddot{C}(0)},
\end{gather}
respectively. The wideband slope, $S_0$, provides the slope of the
spectral efficiency curve $\C(E_b/N_0)$ at zero spectral efficiency
\cite{Verdu}. Therefore, $\left.\frac{E_b}{N_0}\right|_{\C = 0}$ and
$S_0$ constitute a linear approximation of the spectral efficiency
curve in the low-$\tsnr$ regime, i.e.,
\begin{gather}\label{eq:linearapprox}
\C\left(\frac{E_b}{N_0}\right) = \frac{\mathcal{S}_0}{10\log_{10}
2} \left(
\frac{E_b}{N_0}\bigg|_{dB}-\frac{E_b}{N_0}\bigg|_{\C = 0, dB}\right) +
\epsilon
\end{gather}
where $\frac{E_b}{N_0}\Big|_{dB} = 10\log_{10}\frac{E_b}{N_0}$ and
$\epsilon = o\left( \frac{E_b}{N_0}-\frac{E_b}{N_0}\Big|_{\C= 0}\right)$, and characterize the spectral-efficiency/bit-energy tradeoff at low spectral efficiencies. Hence, these quantities enable us to analyze the energy efficiency and investigate the interactions between spectral and energy efficiencies in the low-$\tsnr$ regime. Depending only
on $\dot{C}(0)$ and $\ddot{C}(0)$, the bit energy at zero spectral
efficiency and wideband slope achieved by $M$-ary PSK signals can be
readily obtained by using the formulas in
(\ref{eq:ncfadingfirstderivative}). Note that in the noncoherent
Rician fading channel, the received bit energy is $$
\frac{E_{b,nc}^r}{N_0} = \frac{(|d|^2 + \gamma^2)\tsnr \log
2}{C_{M,nc}(\tsnr)}. $$

\begin{corr:ncfadingbitenergy}
The received bit energy at zero spectral efficiency and wideband
slope achieved by $M$-ary PSK signaling in the noncoherent Rician
fading channel are given by
\begin{align} \label{eq:ncfadingsecondderivative}
\begin{split}
\left.\frac{E_{b,nc}^r}{N_0}\right|_{\C = 0} &=
\left\{
\begin{array}{ll}
\frac{\pi}{2}\left( 1 + \frac{1}{\K}\right)\log 2 & M = 2
\\
\frac{4 \pi}{M^2 \sin^2 \frac{\pi}{M}} \left( 1 +
\frac{1}{\K}\right)\log 2 & M \ge 3
\end{array}\right.
\\ \intertext{and}
S_{0,nc} &= \left\{
\begin{array}{ll}
\frac{3}{\pi-1 + \frac{3\pi}{2\K}}& M = 2
\\
0 & M = 3
\\
\frac{6}{\pi-1 + \frac{3\pi}{\K}} & M = 4
\\
\frac{\frac{M^4}{8\pi^2}\sin^4 \frac{\pi}{M}}{-\psi(M) +
\frac{1}{2\pi \K}M^2 \sin^2\frac{\pi}{M}} & M \ge 5
\end{array} \right.,
\end{split}
\end{align}
respectively, where $\psi(M)$ is given in (\ref{eq:psi}), and $\K =
\frac{|d|^2}{\gamma^2}$ is the Rician factor.
\end{corr:ncfadingbitenergy}

As it will be evident in numerical results, generally the
$\left.\frac{E_{b,nc}^r}{N_0}\right|_{\C = 0}$ is the minimum bit
energy required for reliable transmission when $M \neq 3$. On the
other hand, the minimum bit energy is achieved at a nonzero spectral
efficiency when $M = 3$. Note that this behavior is not exhibited when 3-PSK signals are soft-detected \cite{Kramer}. Hence, this result is tightly linked to correct-detection and error probabilities which are in general functions of the distances in the signal constellation. Note further that at sufficiently low $\tsnr$s, 3-PSK performs worse than 2-PSK (i.e., BPSK), indicating that the decrease in the signal distance from $2\sqrt{\E}$ in 2-PSK to $\sqrt{3\E}$ in 3-PSK has a more dominating effect in the low-$\tsnr$ regime than the increase in the constellation size $M$ from 2 to 3.

%
%

%

\begin{figure}
\begin{center}
\includegraphics[width = \figsize\textwidth]{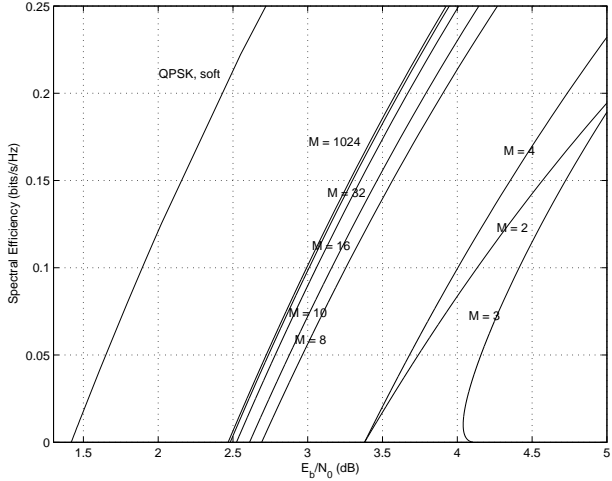}
\caption{Spectral efficiency $\C(E_b/N_0)$ vs. bit energy $E_b/N_0$
for hard-decision detected $M$-ary PSK with $M =
2,3,4,8,10,16,32,1024$ and soft-detected QPSK in the noncoherent
Rician fading channel with Rician factor $\K =
\frac{|d|^2}{\gamma^2} = 1$.} \label{fig:ncfadingbitenergy}
\end{center}
\end{figure}
Figure \ref{fig:ncfadingbitenergy} plots the spectral efficiency
curves as a function of the bit energy for hard-decision detected
PSK with different constellation sizes in the noncoherent Rician
fading channel with Rician factor $\K = 1$.  As observed in this
figure, the information-theoretic analysis conducted in this paper
provides several practical design guidelines. We note that although
hard-decision detected 2-PSK and 4-PSK achieve the same minimum bit
energy of 3.38 dB at zero spectral efficiency, 4-PSK is more
efficient at low but nonzero spectral efficiency values due to its
wideband slope being twice that of 2-PSK. In the range of spectral
efficiency values considered in the figure, 3-PSK performs worse
than both 2-PSK and 4-PSK. 3-PSK achieves its minimum bit energy of
4.039 dB at 0.0101 bits/s/Hz. Operation below this level of spectral
efficiency should be avoided as it only increases the energy
requirements. We further observe that increasing the constellation
size to 8 provides much improvement over 4-PSK. 8-PSK achieves a
minimum bit energy of $2.692$ dB. Note from \eqref{eq:ncfadingsecondderivative} that $\left.\frac{E_{b,nc}^r}{N_0}\right|_{\C = 0}$ is inversely proportional to $M^2 \sin^2 \frac{\pi}{M}$ for $M \ge 3$. Here, we see two competing factors. As $M$ increases, the term $M^2$ increases and tends to decrease the bit energy requirement while the term $\sin^2 \frac{\pi}{M}$ decreases due to a decrease in the minimum distance, which is proportional to $\sin \frac{\pi}{M}$ in $M$-PSK constellation. Hence, when we increase $M$ from 4 to 8, $M^2$ is the dominant term and we note significant gains. As $M$ is further increased, $\sin^2 \frac{\pi}{M}$ acts more strongly to offset the gains from $M^2$ and we see
diminishing returns. For instance, there is little to be gained by
increasing the constellation size more than 32 as 32-PSK achieves a
minimum bit energy of $2.482$ dB and the minimum bit energy as $M
\to \infty$ is $2.468$ dB. 
We find that the wideband slopes of hard-decision detected PSK with
$M$ = 8,10,16,32, and 1024 are 0.571, 0.584, 0.599, 0.607, and
0.609, respectively. The similarity of the wideband slope values is
also
apparent in the figure. 
Note that the wideband slope of 3-PSK, as predicted, is 0.

For comparison, the spectral efficiency of soft-detected QPSK is
also provided in Fig. \ref{fig:ncfadingbitenergy}. It has been shown
in \cite{Gursoy-part2} that under the peak constraint $|x_i|^2 \le
\E$, the bit energy required at zero spectral efficiency and
wideband slope in the noncoherent Rician fading channel with Rician
factor $\K$ are $\left.\frac{E_{b}}{N_0}\right|_{\C = 0} = \left( 1
+ \frac{1}{\K}\right)\log 2$ and $S_0 = \frac{2\K^2}{(1+\K)^2}$,
respectively. It is also proven that soft-detected QPSK is optimally
efficient achieving these values. Note that when $\K = 1$, the bit
energy at zero spectral efficiency is 1.418 dB which is also
observed in Fig. \ref{fig:ncfadingbitenergy}. Note that even as $M
\to \infty$, hard-decision detection presents a loss of  2.468 -
1.418 = 1.05 dB in the minimum bit energy.



%


\subsection{AWGN Channels} \label{subsec:awgnpsk}

Note that the noncoherent Rician fading channel, in which we have
$E\{h_k\} = d$ and $E\{|h_k - d|^2\} = \gamma^2$, specializes to the
AWGN channel if we assume $\gamma^2 = 0$. With this assumption, the
fading coefficients become deterministic, i.e., $h_k = d$, and the
channel model is now $r_k = d s_{x_k} + n_k$ where the channel gain
is $d$. Note also that when we have $\gamma^2 = 0$,
(\ref{eq:nfcondprobr}) becomes the conditional density function of
the output given the input in the AWGN channel. Moreover, the
maximum likelihood decision rule and decision regions for the AWGN
channel are the same as in (\ref{eq:ncdecisionrule}) and
(\ref{eq:ncdecisionregions}), respectively. Assuming further that
$d=1$ leads to the standard unfaded Gaussian channel with unit
channel gain, i.e., the input-output relation becomes $r_k = s_{x_k}
+ n_k$. Based on the above observations, we immediately have the
following Corollary.

\begin{corr:awgnbitenergy}
For the AWGN channel with channel gain $d$, the first and second
derivatives of the PSK capacity at $\tsnr = 0$ are given by the
expressions in (\ref{eq:ncfadingfirstderivative}) if we let
$\gamma^2 = 0$. Furthermore, the bit energy,
$\left.\frac{E_{b}^r}{N_0}\right|_{\C = 0} = \left.\frac{|d|^2\tsnr
\log 2}{C_{M}(\tsnr)}\right|_{\C = 0}$, and wideband slope
expressions are obtained if we let $\gamma^2 = 0$ and hence $\K \to
\infty$ in the formulas in (\ref{eq:ncfadingsecondderivative}).
\end{corr:awgnbitenergy}

\emph{Remark:} We should note that the first derivative of the
capacity of PSK in the AWGN channel has previously been given in
\cite{Pierce_phase} through the bit energy expressions. In addition,
Verd\'u in \cite{Verdu} has provided the second derivative
expression for the special case of $M = 4$. 

\begin{figure}
\begin{center}
\includegraphics[width = \figsize\textwidth]{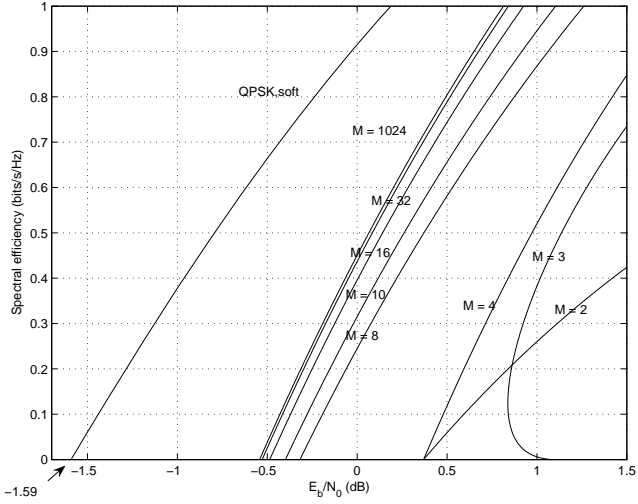}
\caption{Spectral efficiency $\C(E_b/N_0)$ vs. bit energy $E_b/N_0$
for hard-decision detected $M$-ary PSK with $M =
2,3,4,8,10,16,32,1024$ and soft detected QPSK in the AWGN channel.}
\label{fig:awgnbitenergy}
\end{center}
\end{figure}
Fig. \ref{fig:awgnbitenergy} plots the spectral efficiency curves as
a function of the bit energy for hard-decision detected $M$-ary PSK
for various values of $M$ and soft-detected QPSK in the AWGN
channel. Conclusions similar to those given for Fig.
\ref{fig:ncfadingbitenergy} also apply for Fig.
\ref{fig:awgnbitenergy}. The main difference between the figures is
that substantially lower bit energies are needed in the AWGN
channel. For instance, 2- and 4-PSK now achieve a minimum bit energy
of 0.369 dB while 8-PSK attains $-0.318$ dB. As $M \to \infty$, the
minimum bit energy goes to $-0.542$ dB. We note that higher energy
requirements in the noncoherent Rician channel is due to fading and
not knowing the channel.

\subsection{Coherent Fading Channels}

In coherent fading channels, the fading coefficients $\{h_k\}$ are
assumed to be perfectly known at the receiver. We assume that no
such knowledge is available at the transmitter. The only
requirements on the fading coefficients are that their variations
are ergodic and they have finite second moments. Hence, independence
of the random variables $\{h_k\}$ is no longer imposed. Due to the
presence of receiver channel side information (CSI), maximum
likelihood detection is the scaled nearest point detection. In this
case, the average capacity is
\begin{gather} \label{eq:avgcapcoh}
C_{M,c}(\tsnr) = \log M + \sum_{l = 1}^{M} E_h \{P_{l,1,h} \log
P_{l,1,h}\}
\end{gather}
where $$ P_{l,1,h} = P(y = l | x = 1, h) =
\int_{\frac{(2l-3)\pi}{M}}^\frac{(2l-1)\pi}{M} f_{\theta |
s_1,h}(\theta | s_1,h) \,d \theta $$ and $f_{\theta|s_1,h}(\theta|s_1,h)$ is given in \eqref{eq:cfcondprobtheta} on the next page
\begin{figure*}
\begin{align}
f_{\theta|s_1,h}(\theta|s_1,h) = \frac{1}{2 \pi} &\, e^{-|h|^2\tsnr}
+ \sqrt{\frac{|h|^2\tsnr}{\pi}} \cos \theta \,e^{-|h|^2\tsnr \sin^2
\theta} \left( 1 - Q(\sqrt{2|h|^2\tsnr \cos^2 \theta})\right)
\label{eq:cfcondprobtheta}
\end{align}
\hrule
\end{figure*}
with the definition $\tsnr = \E/N_0$. Note that if we assume $\gamma^2 = 0$ and
replace $d$ by the random channel gain $h_k$ in the noncoherent
Rician fading channel, we obtain the model for coherent fading
channels. Hence, similarly as in Section \ref{subsec:awgnpsk},
results for coherent channels can be obtained easily by specializing
those for the noncoherent Rician channel. 
Since we are interested in the average capacity
(\ref{eq:avgcapcoh}), expressions will involve the expected values
of the random gain $h$. Hence, we have the following Corollary to
Theorem \ref{theo:ncfderivatives}.

\begin{corr:cfderivatives} \label{corr:cfderivatives}
The first and second derivatives of $C_{M,c}(\tsnr)$ in nats per
symbol at $\tsnr = 0$ are obtained by assuming in
(\ref{eq:ncfadingfirstderivative}) $\gamma^2 = 0$, replacing $d$ by
$h$, and taking the expectation of the terms that involve $h$. The
resulting expressions are
\begin{align}\label{eq:cfadingfirstderivative}
\begin{split}
\dot{C}_{M,c}(0) &= \left\{
\begin{array}{ll}
\frac{2}{\pi}E\{|h|^2\} & M = 2
\\
\frac{M^2}{4\pi} \sin^2 \frac{\pi}{M} E\{|h|^2\}& M \ge 3
\end{array} \right.,
\\ \intertext{and}
\ddot{C}_{M,c}(0) &= \left\{
\begin{array}{ll}
\frac{8}{3\pi}\left( \frac{1}{\pi} -1\right)E\{|h|^4\} & M = 2
\\
\infty & M = 3
\\
\frac{4}{3\pi}\left( \frac{1}{\pi} -1\right) E\{|h|^4\} & M = 4
\\
\psi(M) E\{|h|^4\} & M \ge 5
\end{array} \right. 
\end{split}
\end{align}
respectively, where $\psi(M)$ is given in (\ref{eq:psi}).
\end{corr:cfderivatives}

\begin{figure}
\begin{center}
\includegraphics[width = \figsize\textwidth]{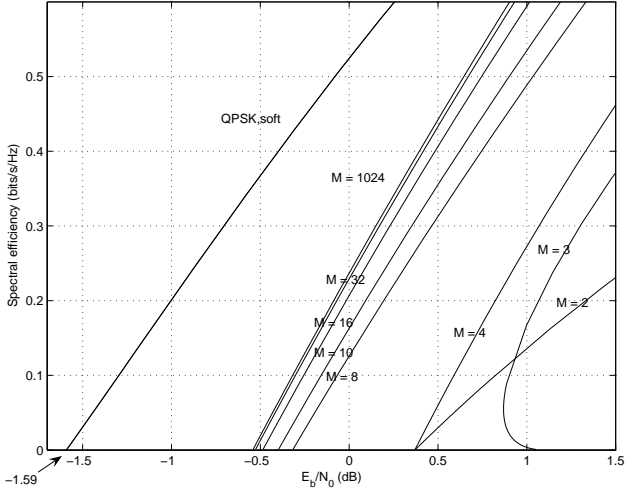}
\caption{Spectral efficiency $\C(E_b/N_0)$ vs. bit energy $E_b/N_0$
for hard-decision detected $M$-ary PSK with $M =
2,3,4,8,10,16,32,1024$ and soft detected QPSK in the coherent
Rayleigh fading channel.} \label{fig:cRayleigh}
\end{center}
\end{figure}

Note that the first and second derivatives of the
capacity at zero $\tsnr$ are essentially equal to the scaled
versions of those obtained in the AWGN channel with $d = 1$. The
scale factors are $E\{|h|^2\}$ and $E\{|h|^4\}$ for the first and
second derivatives, respectively. In the coherent fading case, we
can define the received bit energy as $ \frac{E_{b,c}^r}{N_0} =
\frac{E\{|h|^2\}\tsnr \log 2}{C_M(\tsnr)} $ since $E\{|h|^2\} \tsnr$
is the average received signal-to-noise ratio. It immediately
follows from Corollary \ref{corr:cfderivatives} that $E_b^r / N_0
|_{\C = 0}$ in the coherent fading channel is the same as that in
the AWGN channel. On the other hand, the wideband slope is scaled by
$(E\{|h|^2\})^2/E\{|h|^4\}$. Fig. \ref{fig:cRayleigh} plots the
spectral efficiency curves as a function of bit energy for
hard-decision detected $M$-ary PSK and soft detected QPSK in the
coherent Rayleigh fading channel. Comparison of Fig.
\ref{fig:awgnbitenergy} and Fig. \ref{fig:cRayleigh} reveals that
the bit energy levels required at zero spectral efficiency are
indeed the same for both cases. However, the presence of fading
induces a performance penalty by reducing the wideband slope with a
factor of $E\{|h|^2\}^2/E\{|h|^4\} = 1/2$. Therefore, at low but
nonzero spectral efficiencies, the same bit energy as in the AWGN
channel can be achieved at the cost of reduced spectral efficiency.

\section{Energy Efficiency of Orthogonal Signaling}
\label{sec:fsk}

As discussed in Section \ref{sec:intro}, orthogonal signaling is
optimally energy efficient in the infinite bandwidth regime even if
the receiver performs hard-decision detection. For instance, PPM
with vanishing duty cycle or $M$-ary FSK as $M \to \infty$ achieves
the minimum bit energy of $-1.59$ dB. In this section, we analyze
the non-asymptotic energy
efficiency of orthogonal signaling. 
We consider on-off FSK (OOFSK) modulation in which FSK is combined
with on-off keying (or equivalently PPM) and peakedness is
introduced in both time and frequency. The study of OOFSK modulation
enables us to provide a general unified analysis of orthogonal
signaling as OOFSK can be reduced to OOK and FSK with the
appropriate choice of parameters.

\subsection{OOFSK Modulation}


\subsubsection{AWGN Channels}

In this section, we consider the transmission of OOFSK signals. We
again assume that the received signal is hard-decision detected at
the receiver. In \cite{wang} and \cite{gursoypeaky}, maximum a
posteriori probability (MAP) detection rule for OOFSK modulation  is
identified and the error probability expressions are obtained. We
initially consider the AWGN channel as the results for this channel
will immediately imply similar conclusions for fading channels. The
optimal detection rule in the AWGN channel is given by the
following: $\s_i$ for $i \neq 0$ is detected if
\begin{gather}\label{eq:decisionruleoofsk}
|r_i|^2 > |r_j|^2 \quad \forall j \neq i \quad \text{and} \quad
|r_i|^2
> \tau
\end{gather}
where $\tau = \left\{
\begin{array}{ll}
\frac{\left[I_0^{-1}\left(\xi\right)\right]^2}{4\alpha^2} & \xi \ge
1
\\
0 & \xi < 1
\end{array}\right.$, $ \xi=\frac{M(1-\nu)\,e^{\alpha^2}}{\nu}$, and $\alpha^2 = \frac{\tsnr}{\nu}$. Above, $r_i$ is the $i^{\text{th}}$
component of the received vector $\rr$.
$\s_0$ is detected if $|r_i|^2 < \tau$ $\forall i$. Note that since $\s_0 = (0,\ldots,0)$, detection of $\s_0$ is essentially the detection of no transmission. Note further that the
detection rule in (\ref{eq:decisionruleoofsk}) together with the rule for $\s_0$ can be regarded as energy detection. After detection,
the channel can now be seen as a discrete channel with $M+1$ inputs
and $M+1$ outputs. From the error probability analysis in
\cite{wang} and \cite{gursoypeaky}, we have the
expressions in (\ref{eq:transitionproboofsk2}) through (\ref{eq:transitionproboofsk5}) on the next page for the transition probabilities in the AWGN channel.
\begin{figure*}
\begin{gather}
P_{0,0} = (1-e^{-\tau})^M, \quad \text{and} \quad P_{l,0} =
\frac{1}{M}(1-(1-e^{-\tau})^M) \quad \text{for }  l =
1,2,\ldots,M,\label{eq:transitionproboofsk2}
\\
P_{l,l} =
\sum_{n=0}^{M-1}\frac{(-1)^n}{n+1}\left(\begin{array}{cc}M-1\\n\end{array}\right)
e^{-\frac{n}{n+1}\alpha^2}
Q_1\left(\sqrt{\frac{2}{n+1}}\,\alpha,\sqrt{2(n+1)\tau}\right) \quad
\text{for } l= 1,2,\ldots,M,\label{eq:transitionproboofsk3}
\\
P_{0,l} = (1-e^{-\tau})^{M-1}\left(1-Q_1\left( \sqrt{2}\,\alpha,
\sqrt{2\tau} \right)\right) \quad \text{for } l= 1,2,\ldots,M,
\label{eq:transitionproboofsk4}
\\
P_{l,m} = \frac{1}{M-1}\left(1-P_{m,m}-P_{0,m}\right) \quad
\text{for all } l\neq 0, m\neq 0, \text{and } l \neq m
\label{eq:transitionproboofsk5}
\end{gather}
\hrule
\end{figure*}
In these expressions, $Q_1(\cdot,\cdot)$ is the Marcum $Q$-function \cite{Simon},
and $I_0^{-1}$ is the functional inverse of the zeroth order
modified Bessel function of the first kind. The rates achieved by
the $M$-ary OOFSK modulation with duty cycle $\nu$ and equiprobable
FSK signals is given by (\ref{eq:achievableoofsk}) on the next page.
\begin{figure*}
\begin{align}
I_{M}(\tsnr,\nu) &= H(y) - H(y|x)
\\
&=-\left((1-\nu)P_{0,0} + \nu
P_{0,1} \right) \log \left((1-\nu)P_{0,0} + \nu P_{0,1}
\right)\nonumber
\\
&\,\,\,\,\,\,-M\left( (1-\nu)P_{1,0} + \frac{\nu}{M} P_{1,1} +
\frac{(M-1)\nu}{M} P_{1,2}\right) \log \left( (1-\nu)P_{1,0} +
\frac{\nu}{M} P_{1,1} + \frac{(M-1)\nu}{M} P_{1,2}\right) \nonumber
\\ &\,\,\,\,\,\,+ (1-\nu) \left( P_{0,0} \log P_{0,0} + M P_{1,0} \log
P_{1,0} \right) +\nu \left( P_{0,1} \log P_{0,1} + P_{1,1} \log
P_{1,1} + (M-1) P_{2,1} \log P_{2,1}\right).
\label{eq:achievableoofsk}
\end{align}
\hrule
\end{figure*}
If $M$-ary OOFSK signals have a symbol duration of $T$, the
bandwidth requirement is $\frac{M}{T}$ and the spectral efficiency
is given by $\frac{\frac{I_M(\text{SNR},\nu)}{T}}{\frac{M}{T}} =
\frac{I_M(\text{SNR},\nu)}{M}$ bits/s/Hz.

It is shown in \cite{gursoyoofsk} that in the AWGN channel, the
first derivative of the capacity of soft-detected OOFSK is zero at
$\tsnr = 0$. For the sake of completeness, we provide this result
below.

\begin{cor:finiteParSlope} \label{prop:finitePARslope}
The first derivative of the capacity at zero {\tsnr} achieved by
soft-detected $M$-ary OOFSK signaling with a fixed duty factor $\nu
\in (0, 1]$ over the AWGN channel is zero, and hence the bit energy
required at zero spectral efficiency is infinite.
\end{cor:finiteParSlope}

\emph{Proof}: See \cite{gursoyoofsk}.

Since hard-decision detection does not increase the capacity, we
immediately have the following Corollary to Theorem
\ref{prop:finitePARslope}.

\begin{corr:oofskhard} \label{corr:oofskhard}
The first derivative at zero {\tsnr} of the achievable rates of
hard-decision-detected $M$-ary OOFSK transmission with a fixed duty
factor $\nu \in (0,1]$ over the AWGN channel is zero i.e.,
$\dot{I}_{M}(0,\nu) = 0$, and hence the bit energy required at zero
spectral efficiency is infinite, i.e.,
\begin{gather} \left.\frac{E_{b}}{N_0}\right|_{I = 0} = \frac{\log
2}{\dot{I}_{M}(0,\nu)} = \infty. 
\end{gather}
\end{corr:oofskhard}

On the other hand, we know from \cite{Golay} and \cite{Verdu} that
if the duty cycle $\nu$ vanishes simultaneously with $\tsnr$, the
minimum bit energy of $-1.59$ dB can be achieved. The following
result identifies the rate at which $\nu$ should decrease as $\tsnr$
gets smaller.

\begin{theo:oofskasympt} \label{theo:oofskasympt}
Assume that $\nu = \frac{\tsnr}{(1+\e) \log \frac{1}{\tsnr}}$ for
$\tsnr < 1$ and for some $\e > 0$. Then, we have
\begin{gather}
\lim_{\e \to 0} \lim_{\tsnr \to 0} \frac{I_M(\tsnr,\nu)}{\tsnr} = 1
\intertext{and hence}
\lim_{\e \to 0} \lim_{\tsnr \to 0} \frac{\tsnr
\log 2}{I_M(\tsnr,\nu)} = \log 2 = -1.59 \text{ dB}.
\end{gather}
\end{theo:oofskasympt}

\emph{Proof:} Note that as $\tsnr \to 0$, $\nu \to 0$ and $\alpha^2
= \frac{\tsnr}{\nu} = (1+\e) \log \frac{1}{\tsnr} \to \infty$. It
can also be seen that $\xi \to \infty$ and $\tau \to \infty$ as
$\tsnr$ diminishes. From (\ref{eq:transitionproboofsk2}), we
immediately note that $P_{0.0} \to 1$ and $P_{l,0} \to 0$ for $l =
2,\ldots,M$. In (\ref{eq:transitionproboofsk3}), all the terms in
the summation other than for $n = 0$ vanishes because $\alpha^2 \to
\infty$. Therefore, in order for $P_{l,l}$ for $l=1,\ldots,M$ to
approach 1, we need $Q_1(\sqrt{2}\, \alpha, \sqrt{2\tau}) \to 1$.
Also note that if $Q_1(\sqrt{2}\, \alpha, \sqrt{2\tau}) \to 1$, then
we can observe from (\ref{eq:transitionproboofsk4}) and
(\ref{eq:transitionproboofsk5}) that $P_{0,l} \to 0$ and $P_{l,m}
\to 0$. Hence, eventually all crossover error probabilities will
vanish and correct detection probabilities will be 1.

In \cite{Simon}, it is shown that $Q_1(a, a \zeta) \ge 1 -
\frac{\zeta}{1-\zeta} e^{-\frac{a^2 (1-\zeta)^2}{2}} \quad 0 \le
\zeta < 1. $ From this lower bound we can immediately see that $
\lim_{\tsnr \to 0} Q_1(\sqrt{2}\, \alpha, \sqrt{2\tau}) = 1$ if
$\lim_{\tsnr \to 0} \frac{\tau}{\alpha^2} < 1.$ Note that both
$\alpha^2$ and $\tau$ grow without bound as $\tsnr \to 0$. Recall
that $\tau =
\frac{\left[I_0^{-1}\left(\xi\right)\right]^2}{4\alpha^2}$.
Equivalently, we have $I_0(\sqrt{4\alpha^2 \tau}) = \xi =
\frac{M(1-\nu)\,e^{\alpha^2}}{\nu}$. Using the asymptotic form
$I_0(x) = \frac{1}{\sqrt{2\pi x}}\, e^x +
\mathcal{O}\left(\frac{1}{x^{3/2}}\right)$ \cite{Butkov} for large
$x$, we can easily show that $\lim_{\tsnr \to 0}
\frac{\tau}{\alpha^2} = \left(\frac{1+\e/2}{1+\e}\right)^2 < 1$
$\forall \e > 0$ if $\nu = \frac{\tsnr}{(1+\e) \log
\frac{1}{\tsnr}}$. Therefore, if $\nu$ decays at this rate, the
error probabilities go to zero. It can then be shown that
$\lim_{\tsnr \to 0} \frac{I_M(\tsnr,\nu)}{\tsnr} = \frac{1}{1+\e}$.
Since results hold for any $\e > 0$, letting $\e \to 0$ gives the
desired result. \hfill $\square$

We note that Zheng \emph{et al.} have shown in \cite{Zheng2} that
the low $\tsnr$ capacity of unknown Rayleigh fading channel can be
approached by on-off keying if $\frac{\log \frac{1}{\tsnr}}{\log
\log \frac{1}{\tsnr}} \le \alpha^2 \le \log \frac{1}{\tsnr}$. We see
a similar behavior here when FSK signals are sent over the AWGN
channel and energy detected.

\subsubsection{Fading Channels}

In coherent fading channels where the receiver has perfect knowledge
of the fading coefficients, the transition probabilities are the
same as those in
(\ref{eq:transitionproboofsk2})-(\ref{eq:transitionproboofsk5}) with
the only difference that we now have $\alpha^2 =
\frac{\tsnr}{\nu}|h|^2$. As a result, the achievable rates
$I_{M}(\tsnr,\nu,|h|^2)$ are also dependent on the fading
coefficients and average achievable rates are obtained by finding
the expected value $I_{M,c}(\tsnr,\nu) = E_{|h|^2}
\{I_{M}(\tsnr,\nu,|h|^2)\}$.

In noncoherent Rician fading channels with $E\{h\} = d$ and $E\{|h -
d|^2\} = \gamma^2$, the transition probabilities \cite{wang},
\cite{gursoypeaky} are given by (\ref{eq:transitionproboofsknc2}) through (\ref{eq:transitionproboofsknc5}) on the next page.
\begin{figure*}
\begin{gather}
P_{0,0} = (1-e^{-\tau})^M, \quad \text{and} \quad P_{l,0} =
\frac{1}{M}(1-(1-e^{-\tau})^M) \quad \text{for }  l =
1,2,\ldots,M,\label{eq:transitionproboofsknc2}
\\
\hspace{-.2cm}\!\!\!\!\!\!\!\!\!\!\!\!\!\!\!P_{l,l} =
\sum_{n=0}^{M-1}(-1)^n\left(\!\!\!\begin{array}{cc}M-1\\n\end{array}\!\!\!\right)
\frac{e^{-\frac{n \alpha^2
|d|^2}{n(1+\gamma^2\alpha^2)+1}}}{n(1+\gamma^2\alpha^2 )+1}
Q_1\left(\sqrt{\frac{2 \alpha^2
|d|^2}{(1+\gamma^2\alpha^2)(n(1+\gamma^2\alpha^2)+1)}},\sqrt{\frac{2(n(1+\gamma^2\alpha^2)+1)\tau}{(1+\gamma^2\alpha^2)}}\right)
\, \text{for } l \neq 0, \nonumber \label{eq:transitionproboofsknc3}
\\
P_{0,l} = (1-e^{-\tau})^{M-1}\left(1-Q_1\left( \sqrt{\frac{2\alpha^2
|d|^2}{1+\gamma^2 \alpha^2}}, \sqrt{\frac{2\tau}{1+\gamma^2
\alpha^2}} \right)\right) \quad \text{for } l= 1,2,\ldots,M,
\label{eq:transitionproboofsknc4}
\\
P_{l,m} = \frac{1}{M-1}\left(1-P_{m,m}-P_{0,m}\right) \quad
\text{for all } l\neq 0, m\neq 0, \text{and } l \neq m
\label{eq:transitionproboofsknc5}
\end{gather}
\hrule
\end{figure*}
In these expressions, $$\tau =\left\{
\begin{array}{ll}
 \Phi^{-1}(\xi) & \xi \ge 1
 \\
 0 & \xi < 1
\end{array}\right.$$
where
\begin{gather}
\Phi(x) = e^{\frac{\alpha^2 \gamma^2 x}{1 + \alpha^2 \gamma^2}}
I_0\left( \frac{2\sqrt{x \, \alpha^2 |d|^2}}{1 + \alpha^2
\gamma^2}\right), \text{ and }
\\
\xi = \frac{M(1-\nu)}{\nu} (1+
\alpha^2 \gamma^2) \, e^{\frac{\alpha^2 |d|^2}{1 + \alpha^2
\gamma^2}}.
\end{gather}
The achievable rates, $I_{M,nc}(\tsnr,\nu)$, can be
obtained from (\ref{eq:achievableoofsk}).

Since the presence of fading unknown at the transmitter does not
improve the performance, we readily conclude that the bit energy
requirements in fading channels still grow without bound with
vanishing $\tsnr$.
\begin{corr:fadingfsk} \label{corr:fadingfsk}
The first derivatives at zero $\tsnr$ of the achievable rates
$I_{M,c}(\tsnr,\nu)$ and $I_{M,nc}(\tsnr,\nu)$ are equal to zero,
i.e., $\dot{I}_{M,c}(0,\nu) = \dot{I}_{M,nc}(0,\nu) = 0$. Therefore,
the bit energy required at zero spectral efficiency is infinite in
both coherent and noncoherent fading channels, i.e., $
\left.\frac{E_{b,c}}{N_0}\right|_{I = 0} =
\left.\frac{E_{b,nc}}{N_0}\right|_{I = 0} = \infty. $
\end{corr:fadingfsk}

On the other hand, in noncoherent fading channels, if $|d| \ge 1$,
then following the same steps as in the proof of Theorem
\ref{theo:oofskasympt}, we can show that the minimum bit energy of
$-1.59$ dB is achieved as $\tsnr \to 0$ if $\nu =
\frac{\tsnr}{(1+\e) \log \frac{1}{\tsnr}}$.

\subsection{FSK Modulation}

Recall that if we set $\nu = 1$ in OOFSK modulation, we recover the
regular FSK modulation. Similarly, choosing $\nu = 1$ in the
decision rules and transition probabilities leads to the
corresponding expressions for FSK. For instance, when $\nu = 1$,
$\tau = 0$ in the decision rule (\ref{eq:decisionruleoofsk}) of
OOFSK modulation. Therefore, $\s_i$ is declared as the detected
signal if the $i^{\text{th}}$ component of the received vector $\rr$
has the largest energy, i.e.,
$
|r_i|^2 > |r_j|^2 \quad \forall j \neq i.$ This is the well-known
noncoherent detection of FSK signals. Furthermore, Theorem
\ref{prop:finitePARslope} and Corollaries \ref{corr:oofskhard} and
\ref{corr:fadingfsk} are valid for all $\nu \in (0,1]$ and hence for
$\nu = 1$ as well. Therefore, the same conclusions are automatically
drawn for FSK modulation. Hence, although FSK is energy efficient
asymptotically as $M \to \infty$, operating at very low $\tsnr$
levels with fixed $M$ is extremely energy inefficient as the bit
energy requirement increases without bound with decreasing $\tsnr$.
As a result, the minimum bit energy is achieved at a nonzero
spectral efficiency, the value of which can be found through
numerical analysis. We finally note that when FSK modulation is
considered, the achievable rates are indeed the capacity of FSK
modulation as it is well-known that hard-decision detection capacity
is achieved with equiprobable signals.

\subsection{Numerical Results}

\begin{figure}
\begin{center}
\includegraphics[width = \figsize\textwidth]{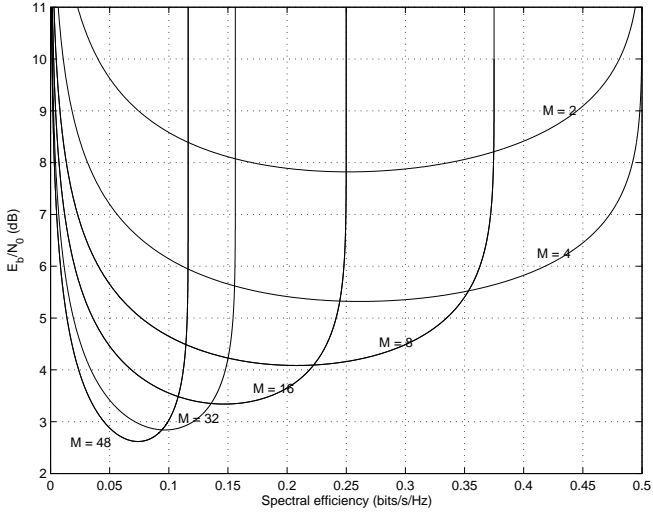}
\caption{Bit energy $E_b/N_0$ vs. Spectral efficiency $\C(E_b/N_0)$
for energy-detected $M$-ary FSK in the AWGN channel.}
\label{fig:fskawgn}
\end{center}
\end{figure}
In this section, we provide numerical results and initially
concentrate on FSK modulation due to its widespread and frequent
use. Fig. \ref{fig:fskawgn} plots the bit energy $E_b/N_0$ curves as
a function of spectral efficiency for $M$-ary FSK in the AWGN
channel for different values of $M$. In all cases, we observe that
the minimum bit energy is achieved at a nonzero spectral efficiency
$\C^*$, and the bit energy requirements increase to infinity as
spectral efficiency decreases to zero. Hence, operation below $\C^*$
should be avoided. Another observation is that the minimum bit
energy and the spectral efficiency value at which the minimum is
achieved decrease with increasing $M$. For instance, when $M = 2$,
the minimum bit energy is 7.821 dB and is achieved at $\C^* = 0.251$
bits/s/Hz. If the value of $M$ is increased to 48, the minimum bit energy
decreases to 2.617 dB and is now attained at $\C^* = 0.074$
bits/s/Hz. Another fact is that as $M$ increases, the minimum bit
energy is achieved at a higher $\tsnr$ value. Indeed, we can show
that
\begin{align}
\lim_{\substack{\e \to 0 \\ M \to \infty}}
\left.\frac{C_M(\tsnr)}{\tsnr}\right|_{\tsnr = (1+\e)\log M}
&=
\lim_{\substack{\e \to 0 \\ M \to \infty}} \frac{C_M((1+\e)\log
M)}{(1+\e)\log M}
\\
&= \lim_{\e \to 0}\frac{1}{1+\e} \,\lim_{M \to
\infty} P_{1,1} = 1. 
\label{eq:c/snrlim}
\end{align}
Hence, if $\tsnr$ grows logarithmically with increasing $M$, the bit
energy $\frac{E_b}{N_0} = \frac{\tsnr \log 2}{C_M(\tsnr)}$
approaches $\log 2 = -1.59$ dB. The proof of (\ref{eq:c/snrlim}) is
omitted because Turin \cite{Turin} has already shown that $-1.59$ dB
is achieved if the signal duration increases as $\log M$, which in
turn increases the $\tsnr$ logarithmically in $M$.

\begin{figure}
\begin{center}
\includegraphics[width = \figsize\textwidth]{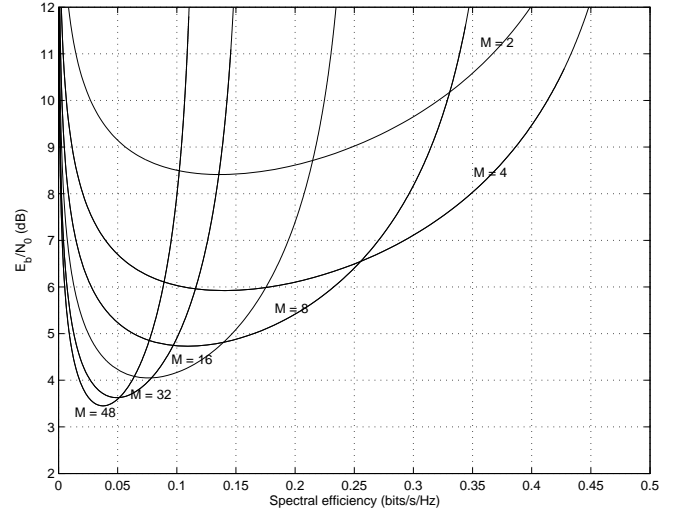}
\caption{Bit energy $E_b/N_0$ vs. Spectral efficiency $\C(E_b/N_0)$
for energy-detected $M$-ary FSK in the coherent Rician fading
channel with Rician factor $\K =1$.} \label{fig:fskcRicianK1}
\end{center}
\end{figure}

\begin{figure}
\begin{center}
\includegraphics[width = \figsize\textwidth]{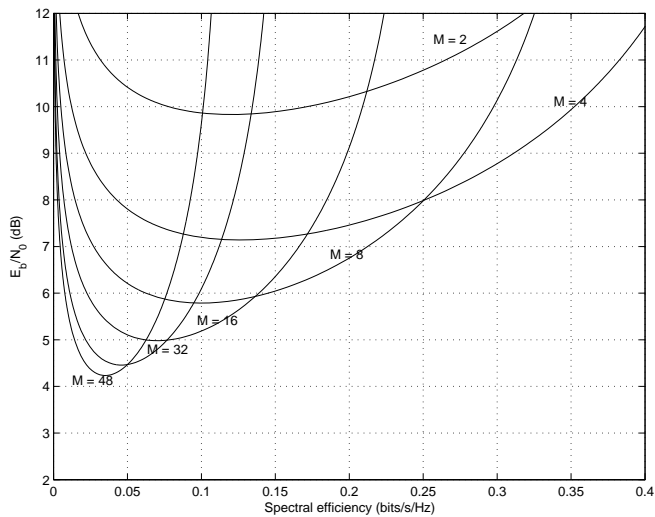}
\caption{Bit energy $E_b/N_0$ vs. Spectral efficiency $\C(E_b/N_0)$
for energy-detected $M$-ary FSK in the noncoherent Rician fading
channel with Rician factor $\K =1$.} \label{fig:fskncRicianK1}
\end{center}
\end{figure}
\begin{figure}
\begin{center}
\includegraphics[width = \figsize\textwidth]{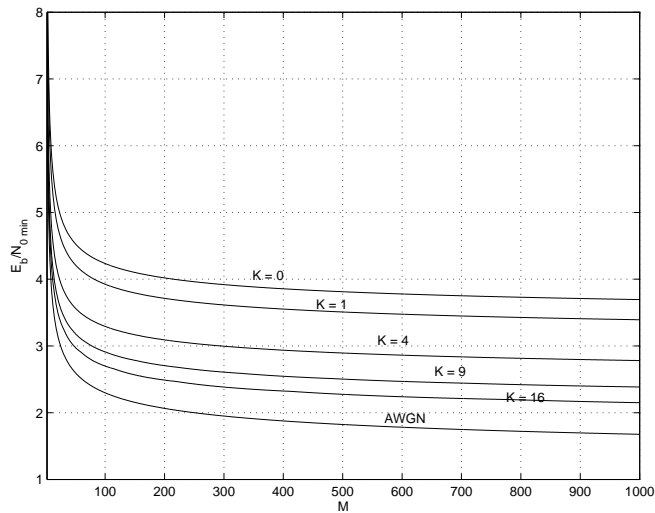}
\caption{Minimum bit energy ${E_b/N_0}_{\min}$ vs. $M$ for $M$-ary
FSK in the AWGN channel and noncoherent Rician fading channels with
Rician factors $\K = 0,1,4,9,16$.} \label{fig:minbitenergy_vs_M2}
\end{center}
\end{figure}
\begin{figure}
\begin{center}
\includegraphics[width = \figsize\textwidth]{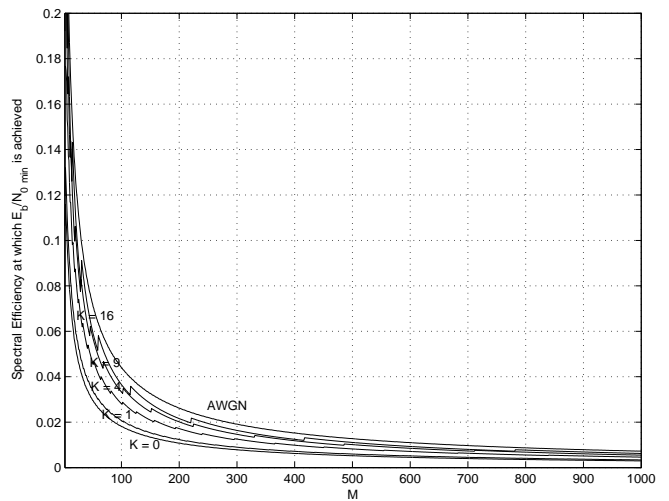}
\caption{Spectral efficiency at which ${E_b/N_0}_{\min}$ is achieved
vs. $M$ for $M$-ary FSK in the AWGN channel and noncoherent Rician
fading channels with Rician factors $\K = 0,1,4,9,16$.}
\label{fig:spectraleff_vs_M}
\end{center}
\end{figure}
\begin{figure}
\begin{center}
\includegraphics[width = \figsize\textwidth]{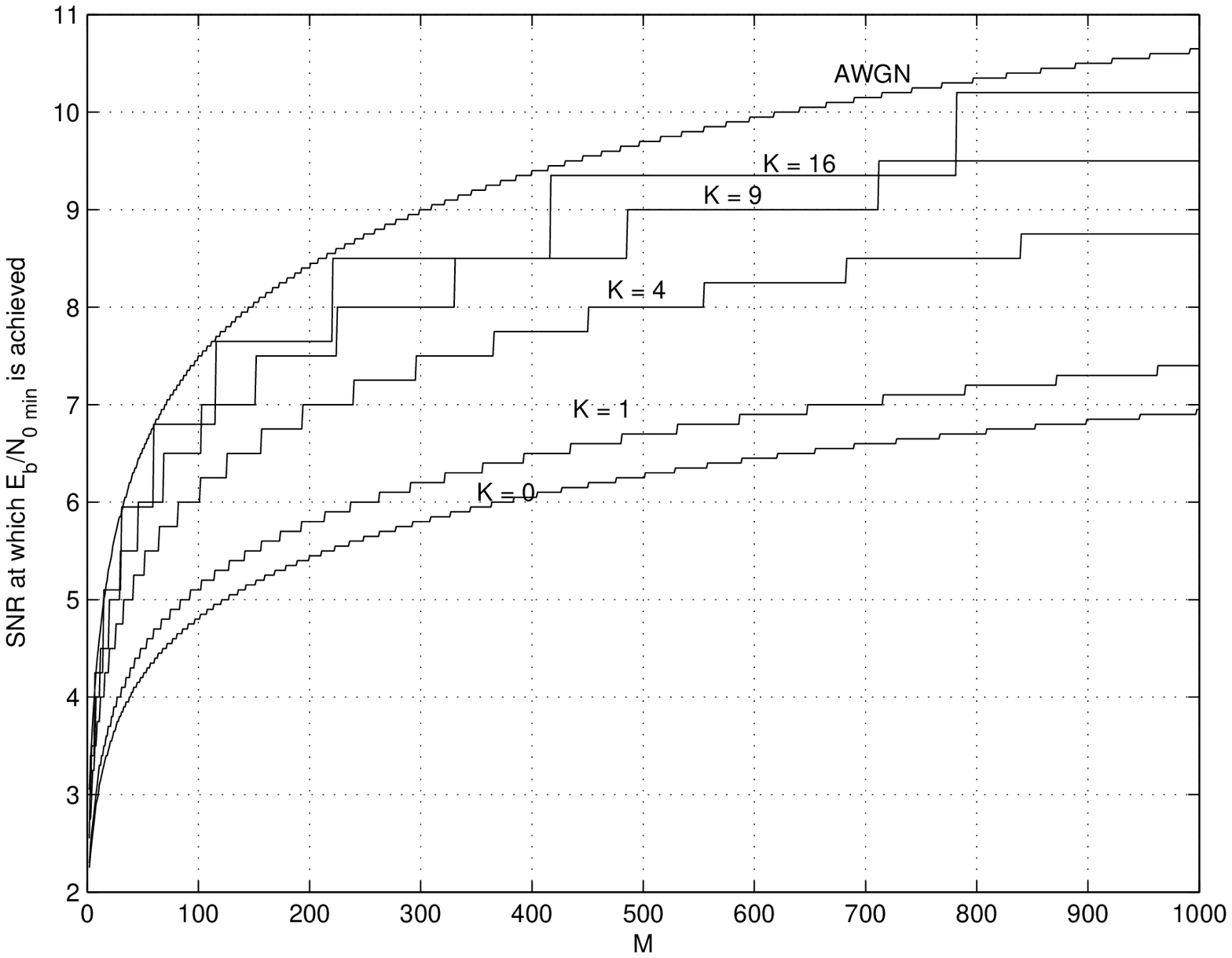}
\caption{$\tsnr$ at which ${E_b/N_0}_{\min}$ is achieved vs. $M$ for
$M$-ary FSK in the AWGN channel and noncoherent Rician fading
channels with Rician factors $\K = 0,1,4,9,16$.}
\label{fig:snr_vs_M}
\end{center}
\end{figure}

Figures \ref{fig:fskcRicianK1} and \ref{fig:fskncRicianK1} plot the
bit energy curves for $M$-ary FSK transmission over coherent and
noncoherent Rician fading channels. As predicted, the bit energy
levels for all values of $M$ increase without bound as the spectral
efficiency decreases to zero. Due to the presence of fading, the
minimum bit energies have increased with respect to those achieved
in the AWGN channel. For instance, when $M = 48$, the minimum bit
energies are now $E_b/N{_0}_{\min} = 3.45$ dB in the coherent Rician
fading channel and $E_b/N{_0}_{\min} = 4.23$ dB in the noncoherent
Rician fading channel. We again observe that the minimum bit energy
decreases with increasing $M$. Fig. \ref{fig:minbitenergy_vs_M2}
provides the minimum bit energy values as a function of $M$ in the
AWGN and noncoherent Rician fading channels with different Rician
factors. In all cases, the minimum bit energy decreases with
increasing $M$. However, Fig. \ref{fig:minbitenergy_vs_M2} indicates
that approaching $-1.59$ dB is very slow and demanding in $M$. In
this figure, we also note the energy penalty due to the presence of
unknown fading. But, as the Rician factor $\K$ increases, the
noncoherent Rician channel approaches to the AWGN channel and so do
the minimum bit energy requirements. Figures
\ref{fig:spectraleff_vs_M} and \ref{fig:snr_vs_M} plot the spectral
efficiencies and average received $\tsnr$ values at which
$E_b/N{_0}_{\min}$ is achieved as a function of $M$. As we have also
observed in Figs. \ref{fig:fskawgn} and \ref{fig:fskncRicianK1}, we see in Fig. \ref{fig:spectraleff_vs_M} that
the spectral efficiency at which $E_b/N{_0}_{\min}$ is achieved decreases with increasing $M$. From Fig. \ref{fig:spectraleff_vs_M}, we further note that the
required spectral efficiencies are lower and hence the bandwidth
requirements are higher in noncoherent fading channels. In Fig.
\ref{fig:snr_vs_M}, we observe that the $\tsnr$ levels at which
$E_b/N{_0}_{\min}$ is achieved increases with increasing $M$. As
predicted by (\ref{eq:c/snrlim}), $\tsnr$ increases logarithmically with $M$
in the AWGN channel. Similar rates of increase are also noted for
the noncoherent fading channel.

\begin{figure}
\begin{center}
\includegraphics[width = \figsize\textwidth]{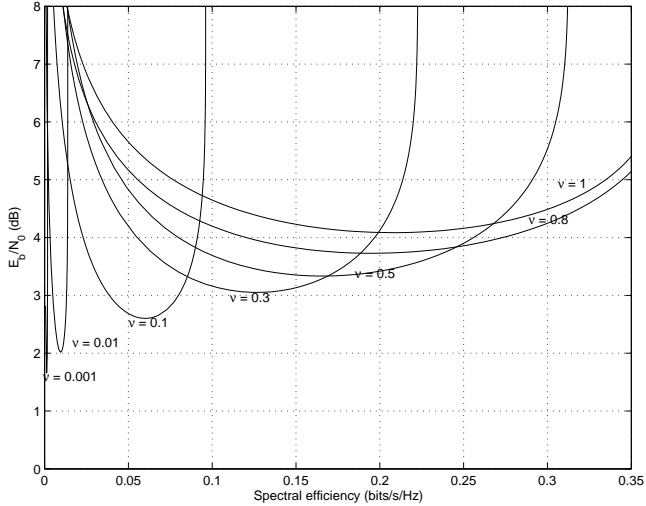}
\caption{Bit energy $E_b/N_0$ vs. Spectral efficiency $\C(E_b/N_0)$
for 8-OOFSK in the AWGN channel. The duty cycle values are $\nu =
1,0.8,0.5,0.3,0,1,0.01$ and 0.001.} \label{fig:oofskawgn}
\end{center}
\end{figure}

\begin{figure}
\begin{center}
\includegraphics[width = \figsize\textwidth]{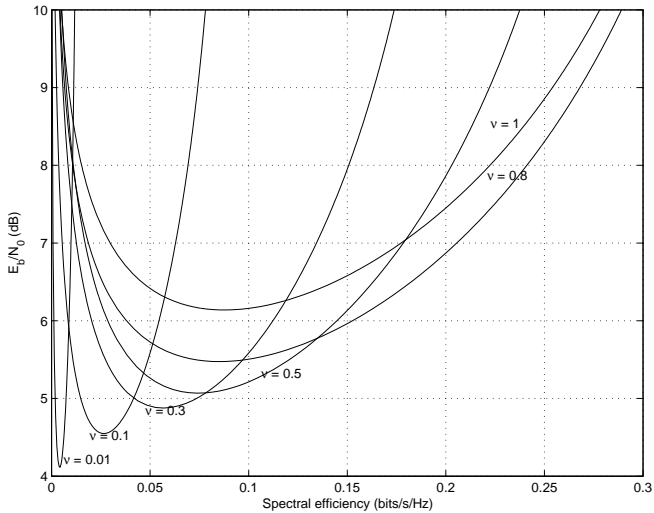}
\caption{Bit energy $E_b/N_0$ vs. Spectral efficiency $\C(E_b/N_0)$
for  8-OOFSK in the noncoherent Rayleigh fading channel. The duty
cycle values are $\nu = 1,0.8,0.5,0.3,0,1,0.01$ and 0.001.}
\label{fig:oofskncRayleigh}
\end{center}
\end{figure}

Figs. \ref{fig:oofskawgn} and \ref{fig:oofskncRayleigh} plot the bit
energies as a function of spectral efficiency of 8-OOFSK with
different duty cycle factors in the AWGN and noncoherent Rayleigh
fading channels. We immediately observe that decreasing the duty
cycle $\nu$ lowers the minimum bit energy. Hence, increasing the
signal peakedness in the time domain improves the energy efficiency.
In the AWGN channel, while regular 8-FSK (8-OOFSK with $\nu =1$) has
$E_b/N{_0}_{\min}= 4.08$ dB, 8-OOFSK with $\nu =0.01$ has
$E_b/N{_0}_{\min} = 2.017$ dB. However, this energy gain is obtained
at the cost of increased peak-to-average ratio. We also note that
unknown fading again induces a energy penalty with respect to that
achieved in the AWGN channel as observed by comparing Figs.
\ref{fig:oofskawgn} and \ref{fig:oofskncRayleigh}.

\section{Conclusion} \label{sec:conclusion}

In this paper, we have analyzed the impact of hard-decision
detection on the energy efficiency of phase modulation and frequency
modulation together with on-off keying. 
We have obtained closed-form expressions for the first and second
derivatives of the $M$-ary PSK capacity in AWGN, coherent fading,
and noncoherent Rician fading channels. Subsequently, we have found
closed-form expressions for the bit energy required at zero spectral
efficiency and wideband slope, and quantified the loss in energy
efficiency incurred by hard-decision detection and channel fading.
The inefficiency of 3-PSK at very low $\tsnr$s is noted. We have
also considered energy detected $M$-ary OOFSK transmission over the
AWGN and fading channels. We have shown that bit energy requirements
grow without bound as $\tsnr$ vanishes for any fixed duty cycle
value. Results are easily specialized to FSK modulation as well.
Through numerical results, we have investigated the value of the
minimum bit energy for different values of $M$ in various channels.
We have shown through numerical results that the minimum bit energy
decreases with decreasing duty cycle and increasing $M$. We have
proved that if the duty cycle decreases as $\frac{\tsnr}{\log
\frac{1}{\tsnr}}$, the minimum bit energy of
$-1.59$ dB can be approached. 

\appendix

\subsection{Proof of Theorem \ref{theo:ncfderivatives}} \label{app:proofthm1}

The main approach is to obtain $\dot{C}_{M,nc}(0)$
and $\ddot{C}_{M,nc}(0)$ by first finding the derivatives of the
transition probabilities $\{P_{l,1}\}$. This can be accomplished by
finding the first and second derivatives of $f_{\theta|s_1}$ with
respect to $\tsnr$. However, the presence of
$\sqrt{\frac{|d|^2\tsnr}{\pi(\gamma^2 \tsnr + 1)}}$ in the second
term of (\ref{eq:nfcondprobtheta}) complicates this approach because
it leads to the result that
$\left.\frac{df_{\theta|s_1}}{d\ssnr}\right|_{\ssnr = 0} = \infty.$
In order to circumvent this problem, we define the new variable $a =
\sqrt{\tsnr}$ and consider the conditional density expression in (\ref{eq:nfcondprobthetaa}) on the next page.
\begin{figure*}
\begin{align}
f_{\theta|s_1}(\theta|s_1) = \frac{1}{2 \pi} \,
e^{-\frac{|d|^2a^2}{\gamma^2 a^2 + 1}} +
\frac{|d|a}{\sqrt{\pi(\gamma^2 a^2 + 1)}} \cos \theta
\,e^{-\frac{|d|^2a^2}{\gamma^2 a^2 + 1} \sin^2 \theta} \left( 1 -
Q\left(\sqrt{2\frac{|d|^2a^2}{\gamma^2 a^2 + 1} \cos^2
\theta}\right)\right). \label{eq:nfcondprobthetaa}
\end{align}
\hrule
\end{figure*}
Now, the derivative expressions in (\ref{eq:derivatives1}) and (\ref{eq:derivatives2}) on the next page evaluated at $a= 0$ can easily be
verified.
\begin{figure*}
\begin{gather}
f_{\theta|s_1}(\theta|s_1)|_{a = 0} =
\frac{1}{2\pi}, \qquad \left.\frac{df_{\theta|s_1}}{da}\right|_{a =
0} = \frac{|d|\cos \theta}{2 \sqrt{\pi}}, \qquad
\left.\frac{d^2f_{\theta|s_1}}{da^2}\right|_{a = 0} =
\frac{|d|^2\cos 2 \theta}{\pi} \label{eq:derivatives1}
\\
\left.\frac{df^3_{\theta|s_1}}{da^3}\right|_{a = 0}
= -\frac{3|d|\gamma^2 \cos \theta}{2\sqrt{\pi}} -\frac{3 |d|^3 \cos
\theta \sin^2 \theta}{\sqrt{\pi}}, \qquad
\left.\frac{df^4_{\theta|s_1}}{da^4}\right|_{a = 0} =
-\frac{12|d|^2\gamma^2\cos 2\theta}{\pi} + \frac{6 |d|^4 \cos^2
2\theta}{\pi} - \frac{8 |d|^4 \cos^4 \theta}{\pi}.\label{eq:derivatives2}
\end{gather}
\hrule
\end{figure*}
Using the derivatives of $P_{l,1}$ and performing several algebraic
operations, we arrive to the following Taylor expansion of
$C_{M,nc}(a)$ at $a = 0$:
\begin{align}
\!\!\!\!\!\!\!C_{M,nc}(a) &= \phi_1(M) \,a^2 + \phi_2(M) \, a^3 +
\phi_3(M) \, a^4 + o(a^4) \label{eq:expansion1}
\\
&=\phi_1(M) \tsnr + \phi_2(M) \tsnr^{3/2} + \phi_3(M) \tsnr^2 +
o(\tsnr^2) \label{eq:expansion2}
\end{align}
where (\ref{eq:expansion2}) follows due to the fact that $a =
\sqrt{\tsnr}$. In the above expansion, $\phi_1(M)$, $\phi_2(M)$, and $\phi_3(M)$ are given by (\ref{eq:phi1})--(\ref{eq:phi3}) on the next page.
\begin{figure*}
\begin{align}
\phi_1(M) &= \frac{M|d|^2}{2\pi} \sin^2 \frac{\pi}{M}
\sum_{i=1}^M \cos^2 \frac{2\pi i}{M}, \label{eq:phi1}
\\
\phi_2(M) &= \frac{M
|d|^3}{\pi\sqrt{\pi}} \left(\sin \frac{\pi}{M} \sin \frac{2\pi}{M} -
\frac{M}{6} \sin^3 \frac{\pi}{M}\right) \sum_{i=1}^M \cos^3
\frac{2\pi i}{M}, \label{eq:phi2}
\\
\phi_3(M) &= -\frac{M^2|d|^4}{16\pi} \sin^2
\frac{2\pi}{M} + \frac{M|d|^4(\pi+2)}{16\pi^2}\sin^2\frac{2\pi}{M}
\sum_{i=1}^M \cos^2 \frac{4\pi i}{M} \nonumber
\\
&\hspace{.5cm}+|d|^4\left( \left(\frac{M^3}{12\pi^2} -
\frac{M}{3\pi}\right)\sin^4 \frac{\pi}{M} - \frac{M^2}{2\pi^2}
\sin^2 \frac{\pi}{M} \sin \frac{2\pi}{M}\right) \sum_{i=1}^M \cos^4
\frac{2\pi i}{M} +\frac{M^2}{4 \pi^2} \sin^2 \frac{\pi}{M} \sin
\frac{2\pi}{M} \sum_{i=1}^M \cos^2 \frac{2\pi i}{M} \nonumber
\\
&\hspace{.5cm}-\frac{|d|^2 \gamma^2}{2\pi} \, M \sin^2 \frac{\pi}{M} \sum_{i =
1}^M \cos^2 \frac{2 \pi i}{M}.\label{eq:phi3}
\end{align}
\hrule
\end{figure*}
We immediately conclude from (\ref{eq:expansion2}) that
$\dot{C}_{M,nc}(0) = \phi_1(M)$. Note that the expansion includes
the term $\tsnr^{3/2}$ which implies that $\ddot{C}_{M,nc}(0) =
\pm\infty$ for all $M$. However, it can be easily seen that
$\phi_2(M) = 0$ for all $M \neq 3$, and at $M = 3$, $\phi_2(3) =
0.1718|d|^3$. Therefore, while $\ddot{C}_{3,nc}(0) = \infty$,
$\ddot{C}_{M,nc}(0) = 2\phi_3(M)$ for $M \neq 3$. Further algebraic
steps and simplification yield (\ref{eq:ncfadingfirstderivative}).
\hfill $\square$


\end{document}